\documentclass[aps,prb,twocolumn,amssymb,amsmath,superscriptaddress]{revtex4-1}
\usepackage{bm}
\usepackage{times}
\usepackage{graphicx}
\usepackage{color}
\usepackage{dcolumn}
\usepackage[colorlinks=true, letterpaper=true, pdfstartview=FitV, linkcolor=blue, citecolor=blue, urlcolor=blue]{hyperref}
\usepackage{appendix}

\begin{document}
\title{ General method for calculating transport properties of disordered mesoscopic systems \\
based on the nonequilibrium Green's function formalism }
\author{Gaoyang Li$^\S$}
\affiliation{Graduate School of China Academy of Engineering Physics, Beijing 100193, China}
\affiliation{College of Physics and Optoelectronic Engineering, Shenzhen University, Shenzhen 518060, China}

\author{MiaoMiao Wei$^\S$}
\affiliation{College of Physics and Optoelectronic Engineering, Shenzhen University, Shenzhen 518060, China}

\author{Fuming Xu}
\affiliation{College of Physics and Optoelectronic Engineering, Shenzhen University, Shenzhen 518060, China}

\author{Jian Wang}
\email[]{jianwang@hku.hk}
\affiliation{College of Physics and Optoelectronic Engineering, Shenzhen University, Shenzhen 518060, China}
\affiliation{Department of Physics, The University of Hong Kong, Pokfulam Road, Hong Kong, China}

\begin{abstract}
Disorder scattering plays important roles in quantum transport as well as various Hall effects, including the second-order nonlinear Hall effect induced by Berry curvature dipole. Calculation of disorder-averaged transport properties usually requires substantial computational resources, especially for higher-order effects. Existing methods are either limited by approximation conditions or constrained by numerical stability, making it difficult to conveniently obtain average physical quantities over a wide range of disorder strength. In this work, we develop a general method for noninteracting system to obtain analytical expressions of disorder averages in finite orders of disorder strength. This method utilizes the Dyson equation to expand physical quantities expressed in terms of the Green's functions into series of disorder-averaged matrices, and the only approximation involved is the truncation of the Dyson equation. Therefore, this method not only avoids the brute force calculation of disorder samples, but also widely applies to different model systems, types of disorder, and the number of Green's functions in the expressions. We demonstrate the applicability of this general method by calculating averages of the linear conductance of a two-terminal system, the spin Hall conductance and the second-order nonlinear conductance of four-terminal Hall setups. It is found that truncation at the fourth order of disorder strength provides a reasonable accuracy and a convenient Pad\'{e} treatment effectively extends its applicable range. Numerical results also confirms disorder enhancement of the second-order nonlinear Hall current in four-terminal systems. Moreover, more accurate predictions for a broader range of disorder strength can be achieved by including higher-order terms in a similar manner.
\end{abstract}

\maketitle

\section{INTRODUCTION}

Disorder effect has been extensively studied for decades, especially since the concept of Anderson localization was introduced~\cite{Anderson}. Investigation on disordered systems is not only a fundamental issue in condensed matter physics, but also holds significant implications in application areas of electronics~\cite{Lee} and spintronics~\cite{Reinhold,Chadov,Rocha,Schedin}, since the performance of electronic devices is severely influenced by randomness arising from impurities, defects, and roughness. Disorder effect is particularly pronounced in mesoscopic transport regime, where prominent quantum interference gives rise to universal fluctuation in charge conductance~\cite{Altshuler,Lee1,Lee2,Beenakker97,Qiao10} and spin Hall conductance~\cite{Ren,Qiao08}. In topological systems, disorder scattering can induce the topological Anderson insulator phase~\cite{Li09,Beenakker09,HMGuo10,Xing11}. In addition, disorder scattering mechanisms such as skew scattering and side jump, play crucial roles in anomalous Hall effect~\cite{Nagaosa10} and spin Hall effect~\cite{Sinova15}. Recently, disorder enhancement of the second-order Hall effect induced by Berry curvature dipole has been reported~\cite{Du2019,Du2021,Yao2022}. Similar enhancement behavior is also discovered for quantum third-order Hall effect in a four-terminal system~\cite{Wei2022}. To summarize, quantitative evaluation of disorder-averaged transport properties including Hall responses in mesoscopic systems, is of great importance.

In quantum transport studies, the Landauer-B\"ttiker (LB) formalism~\cite{Landauer,Buttiker2} provides a framework to express conductance in terms of transmission probabilities, which can be determined using methods like the scattering matrix and Green's function (GF) techniques~\cite{Datta}. The GF method has led to significant advancements in the study of disordered transport, giving rise to a variety of approaches such as the recursive Green's function~\cite{Thouless,Drouvelis,MacKinnon,Baranger,Sols,Kazymyrenko,Ferry,note1}, the coherent potential approximation (CPA)\cite{Elliott,Velicky,Taylor,Soven,Cheng2,Li}, and CPA with nonequilibrium vertex correction (CPA-NVC)\cite{Ke1,Ke2,Yan,Yan2,Cheng,Zhai,Kalitsov,Zhu}. Additionally, full counting statistics within CPA (FCS-CPA)\cite{Fu,Zhang} has emerged as an important tool. A more recent advancement, nonequilibrium dynamical cluster theory, extends beyond CPA by incorporating nonlocal disorder correlations and accounting for the short-range order of disorder scattering\cite{Zhang3,Zhang4}. The Kubo formalism, expressed using Chebyshev polynomials, also proves to be an effective approach for studying disordered systems~\cite{Castro,Castro2}.

To obtain converged average transport properties in disordered systems, a natural way is averaging the physical quantity over a large ensemble of disordered samples, which is known as the brute force (BF) calculation. Typically, this requires a considerable amount of computational resources. Recursive Green's function method reduces the time needed to compute individual disorder sample, but still requires calculations for a large ensemble. CPA introduces the single site approximation (SSA) and neglects multiple scattering processes of electrons at the lattice sites, which can directly obtain the average of a single retarded GF through self-consistent solution of the effective disorder potential. However, this method can only provide reliable results when the disorder strength is relatively weak, and the self-consistent process is usually unstable and difficult to converge~\cite{Zhang2} when disorder increases. CPA-NVC considers the effects of multiple scattering and can calculate the average of a pair of GFs, from which one can calculate the average conductance and current~\cite{Ke1,Ke2,Kalitsov,Zhu}. The CPA-NVC method has also been generalized to handle the disorder average of two-particle NEGF correlators, such as fluctuations and shot noise~\cite{Yan,Yan2}. FCS-CPA is the generalization of CPA~\cite{Fu}. It only requires solving a generating function to obtain the averages of any number of GFs and all orders of their cumulants. Recently, FCS-CPA has been applied to study the statistics of spin transport through nonmagnetic metal/ferromagnetic insulator hybrid system~\cite{Li}. Although FCS-CPA is a powerful tool for calculating disorder-related properties, it still inherits the drawback of CPA. Therefore, there is an urgent need for a general method for quantum transport through disordered devices that avoids massive BF calculations and strong approximations.

In this paper, we develop such a method. Based on the nonequilibrium Green's function (NEGF) formalism, we combine it with the Dyson equation to expand disorder-averaged physical quantities into series of disorder matrices. By incorporating the specific properties of disorder matrices, we obtain the analytical expressions of disorder-averaged quantities as a function of the disorder strength. This method circumvents the limitations of the aforementioned approaches, and is not restricted to specific model systems, types of disorder, and the number of GFs in the expressions of physical quantities. The only approximation in this method is the truncation of the Dyson equation, and its accuracy can always be improved by including higher-order terms in a systematic way, which can be easily realized in codes. Therefore, it exhibits broad applicability. As a demonstration, we
calculate averages of the linear conductance of a two-terminal system, the spin Hall conductance and the second-order nonlinear conductance of four-terminal Hall setups. We find that the truncation in the fourth order of disorder strength already gives good estimations in a wide range of disorder, which can be effectively extended by a simple Pad\'{e} treatment. Our method presents the analytical expressions of disorder averages as a function of the disorder strength, which can also provide guidance for similar research on disorder-averaged properties.

The rest of this paper is organized as follows. In Sec.~\ref{sec:theory}, we illustrate the method by showing how to compute the average linear conductance in a two-terminal system and the average second-order conductance in a four-terminal system. In Sec.~\ref{sec:res}, we calculate the average linear and second-order nonlinear transport properties for three different model systems, where the method is expanded to the second and fourth orders of disorder strength and then compared with BF results. A conclusion is drawn in Sec.~\ref{sec:conclusion}.

\section{THEORETICAL FORMALISM}\label{sec:theory}
In this section, we delineate the theoretical formalism employing nonequilibrium Green's functions. We consider a system consisting of a central region connected to multiple leads. The central region is disordered due to impurities and defects etc. The Hamiltonian of the disordered system is generally expressed as follows:
\begin{equation}\label{H}
H = H_C + V + H_{\rm leads} + H_T,
\end{equation}
where $H_C$ is the Hamiltonian of the central region, $V$ represents on-site disorder in the central region. $H_{\rm leads}$ characterizes the Hamiltonians of the leads and $H_T$ denotes the coupling between the central region and the leads.

In the regime of weak disorder, the influences of such disorder can be treated perturbatively. The retarded Green's function for the unperturbed Hamiltonian $H_0=H_C + H_{\rm leads} + H_T$ is given by
\begin{equation}
g^r = \frac{1}{E-H_0-\Sigma^r},
\end{equation}
where $\Sigma^r=\sum_{\alpha}\Sigma_{\alpha}^{r}$ is the total self-energy of all leads. The self-energy associated with lead $\alpha$ is calculated from the transfer-matrix method~\cite{Lee3,Lee4}. Upon turning on the disorder matrix $V$, we can formulate the Green's function of the perturbed Hamiltonian in Eq.~(\ref{H}) using the Dyson equation as
\begin{equation}\label{eq:Dyson}
\begin{split}
G^r &= g^r + g^rVG^r\\
& = g^r + g^rVg^r + g^rVg^rVg^r + \cdots.
\end{split}
\end{equation}
Consequently, physical quantities, such as ${\cal{O}}$, can be expressed in terms of disorder by expanding the Green's functions $G^{r,a}$ with respect to $V$,
\begin{equation}
{\cal{O}} = f(G^r) = f_0(g^r) + f_1(g^r, V) + f_2(g^r, V^2) + \cdots,
\end{equation}
where $\cal{O}$ represents an arbitrary physical quantity and functions $f_{n}$ denote the expanded terms in the $n$th-order of $V$. The average $\langle {\cal{O}} \rangle$ is thus given by summation of each averaged terms above. By using the properties of disorder $V$, it is always feasible to get an analytical expression for each term. Then, the quantity $\langle {\cal{O}} \rangle$ is expressed in orders of disorder parameters. For the convenience of demonstration, we assume an Anderson-type disorder, a random on-site potential energy drawn from an uniform distribution in the range of $\left[-W/2,W/2\right]$, where $W$ is the disorder strength. Other types of disorder can be handled similarly by employing the respective probability density functions~\cite{note2}. Thus we obtain
\begin{equation}
\langle {\cal{O}} \rangle = a_0 + a_1 W + a_2W^2 + a_3W^3 + \cdots,
\end{equation}
in which, $a_n$ are coefficients independent of disorder and are expressed in terms of Green's functions.

In the following subsections, we will provide two examples illustrating the application of our formalism in different models. We emphasize that our formalism is independent of specific model Hamiltonians and disorder types. The sole approximation involved in this formalism lies in the truncation of the Dyson equation, a procedure that can always be refined by incorporating higher-order terms.

\subsection{Example: average conductance in a two-terminal system}
In this subsection, we take the average conductance in a two-terminal normal metal (NM) system as an example to illustrate our formalism. The conductance in such a system is formulated as $T= 2e^2/h ~ {\rm Tr}[\Gamma_L G^a \Gamma_R G^r]$~\cite{Buttiker}, where $\Gamma_{L,R}$ are linewidth functions associated with the left and right leads. $G^{r,a}$ are Green's functions of the central scattering region. For simplicity, we set $2e^2/h = 1$ in the following. In the presence of disorder, the average conductance is given by
\begin{equation}\label{eq:<TT>}
\langle T\rangle = {\rm Tr}[ \Gamma_L\langle G^r \Gamma_R G^a\rangle],
\end{equation}
in which, $\langle \cdots \rangle$ denotes the average of different disorder samples under a specific disorder strength. Expanding $G^{r,a}$ using Eq.~(\ref{eq:Dyson}) and keeping terms up to the fourth order of $V$, we have
\begin{equation}\label{eq:<T>}
\langle T\rangle = T^{(0)} + \langle T^{(2)}\rangle + \langle T^{(4)}\rangle + O(V^{6}),
\end{equation}
where we have used the fact that for Anderson disorder, $\langle V^n\rangle=0$ for odd $n$, then all odd terms disappear. The zeroth-order term $T^{(0)}$ is disorder-independent
\begin{equation}\label{eq:T0}
T^{(0)}={\rm Tr}[\Gamma_Lg^r \Gamma_R g^a].
\end{equation}
The second-order term involving disorder $V$ reads
\begin{equation}
\begin{split}
\langle T^{(2)}\rangle = {\rm Tr}&[D_L\langle V g^r V\rangle g^r \Gamma_R + D_L\langle V g^r \Gamma_R g^a V\rangle \\
&+ D_L \Gamma_R g^a\langle V g^a V\rangle].
\end{split}
\end{equation}
Each term in $\langle T^{(2)}\rangle$ can be calculated in a similar manner. For convenience, the matrix production $D_L=g^a \Gamma_L g^r$ is adopted. After regrouping, we can denote each term in $\langle T^{(2)}\rangle$ as
\[ \langle T_i^{(2)}\rangle = {\rm Tr} [A\langle VBV\rangle],\]
where $A$ and $B$ are matrix productions independent of disorder. Given that the disorder manifests as an on-site potential within the central region, the disorder matrix $V$ is diagonal. Therefore, the matrix element of the disorder average $X^{(2)}=\langle VBV\rangle$ can be expressed as
\begin{equation}
\begin{split}
X_{ij}^{(2)}={[\langle VBV\rangle]}_{ij} &= B_{ij}\langle V_{ii}V_{jj}\rangle.
\end{split}
\end{equation}
In the case of a single-band model, this simplifies to~\cite{note2}
\begin{equation}\label{eq:V2}
\langle V_{ii}V_{jj}\rangle = \frac{1}{W}\int_{-W/2}^{W/2} V^2 dV\delta_{ij} = \frac{W^2}{12}\delta_{ij}.
\end{equation}
Note that in the case of multi-band models, the Kronecker $\delta$ function should be replaced by the function $\delta'_{i,j}$ defined as
\begin{equation}\label{eq:delta}
\delta'_{i,j}=
\begin{cases}
1,& \text{if $i$ and $j$ belong to the same site,} \\
0, & \text{otherwise.}
\end{cases}
\end{equation}
For the single band model considered here, the second-order term of the average conductance is expressed as
\begin{equation}\label{eq:<T2>}
\begin{split}
\langle T^{(2)}\rangle =& \frac{W^2}{12}\{{\rm Tr}[D_L \tilde{X}_1^{(2)} g^a \Gamma_R ] + {\rm Tr}[D_L \tilde{X}_2^{{(2)}}] \\
&+ {\rm Tr}\Big[D_L \Gamma_R g^a [\tilde{X}_1^{(2)}]^\dag \Big]\},
\end{split}
\end{equation}
with the transformed matrices defined as
\begin{eqnarray}
\begin{aligned}
\big[\tilde{X}_1^{(2)}\big]_{ii} &= \big[g^r\big]_{ii},\\
\big[\tilde{X}_2^{(2)}\big]_{ii} &= \big[g^r \Gamma_R g^a\big]_{ii},
\end{aligned}
\end{eqnarray}
which are diagonal matrices.

In general, the second-order approximation in $V$ only provides accurate average at small disorder strength. To achieve reliable prediction for stronger disorder, it is necessary to include higher-order terms. Here we incorporate the fourth-order terms. Expanding the Green's functions in the Dyson equation and retaining the fourth-order terms, we obtain
\begin{equation}
\begin{split}
\langle T^{(4)}\rangle =
&  {\rm Tr}[D_L \Gamma_R g^a\langle V g^a V g^a V g^a V\rangle]\\
&+ {\rm Tr}[D_L\langle V g^r \Gamma_R g^a V g^a V g^a V\rangle]\\
&+ \frac{1}{2}{\rm Tr}[D_L\langle V g^r V g^r \Gamma_R g^a V g^a V\rangle] + c.c
\end{split}
\end{equation}
Notice that each term in $\langle T^{(4)}\rangle$ involves four disorder matrices $V$. By consolidating matrix productions that are independent of disorder $V$ into $A, B, C$ and $D$, we can write each term as
\begin{equation}
\langle T_i^{(4)}\rangle = {\rm Tr}[A\langle V B V C V D V\rangle].
\end{equation}
Defining the matrix $X^{(4)}=\langle V B V C V D V\rangle$, we have
\begin{equation}
X_{ij}^{(4)} = \sum_{kl} B_{i k} C_{k l} D_{l j} \epsilon_{ijkl},
\end{equation}
where the factor $\epsilon_{ijkl}$ is given by
\begin{equation}\label{eq:X4}
\epsilon_{ijkl}= \begin{cases}
\langle V_{ii}V_{jj}\rangle\langle V_{k k}V_{ll}\rangle + \langle V_{ii}^4\rangle\delta_{ij}\delta_{kl}\delta_{jk} ,& i=j, \\
\langle V_{ii}V_{kk}\rangle\langle V_{ll}V_{jj}\rangle + \langle V_{ii}V_{ll}\rangle\langle V_{kk}V_{jj}\rangle, & i\neq j.
\end{cases}
\end{equation}
The second-order term $\langle V_{ii}V_{jj}\rangle$ has been given in Eq.~(\ref{eq:V2}). Similarly, the fourth-order term is found to be~\cite{note2}
\begin{equation}\label{eq:V4}
\langle V_{ii}^4\rangle = \frac{W^4}{80}.
\end{equation}
Again, for multi-band models, the Kronecker $\delta$ function should be replaced by Eq.~(\ref{eq:delta}). Therefore, we get the analytical expression for the fourth-order terms
\begin{equation}\label{eq:<T4>}
\begin{split}
\langle T^{(4)}\rangle =
W^4\{
&  {\rm Tr}[D_L \Gamma_R g^a \tilde{X}_1^{(4)}] + {\rm Tr}[D_L \tilde{X}_2^{(4)} ]\\
&+ {\rm Tr}[D_L g^r \tilde{X}_3^{(4)}] + {\rm Tr}[D_L \tilde{X}_4^{(4)}]\\
&+ {\rm Tr}[D_L \tilde{X}_5^{(4)} g^r \Gamma_R]\},
\end{split}
\end{equation}
where the disorder-averaged matrices $\tilde{X}_i^{(4)}$ can be extracted from Eq.~(\ref{eq:X4}). For example, the first matrix $\tilde{X}_1^{(4)}$ and the second matrix $\tilde{X}_2^{(4)}$ are given by
\begin{equation}
[\tilde{X}_{m}^{(4)}]_{ij}=
\begin{cases}
\sum_k\frac{1}{144} [B_m]_{ik} [g^a]_{kk} [g^a]_{ki} + \frac{1}{80} [B_m]_{ii} [g^{a}]_{ii}^2,& i=j, \\
\frac{1}{144} ([B_m]_{ii} [g^a]_{ij} [g^a]_{jj} + [B_m]_{ij} [g^a]_{ji} [g^a]_{ij}), & i\neq j,
\end{cases}\nonumber
\end{equation}
with $m=1, 2$. Here the matrices $B_1=g^a$ and $B_2=g^r \Gamma_R g^a$ for terms $\tilde{X}_1^{(4)}$ and $\tilde{X}_2^{(4)}$, respectively.

Combining Eqs.~(\ref{eq:<T>}), (\ref{eq:T0}), (\ref{eq:<T2>}) and (\ref{eq:<T4>}), we get the analytical expression of $\langle T\rangle$ up to the fourth order in $W$,
\begin{equation}\label{eq:Tw}
\langle T\rangle = a_0 + a_2W^2 + a_4W^4 + O(W^6),
\end{equation}
with coefficients $a_n$ expressed in terms of Green's functions. Solving the matrix products in $a_n$ gives prediction on $\langle T \rangle$, which avoids the time-consuming brute force calculation.

\begin{figure}[tbp]
\includegraphics[width=\columnwidth]{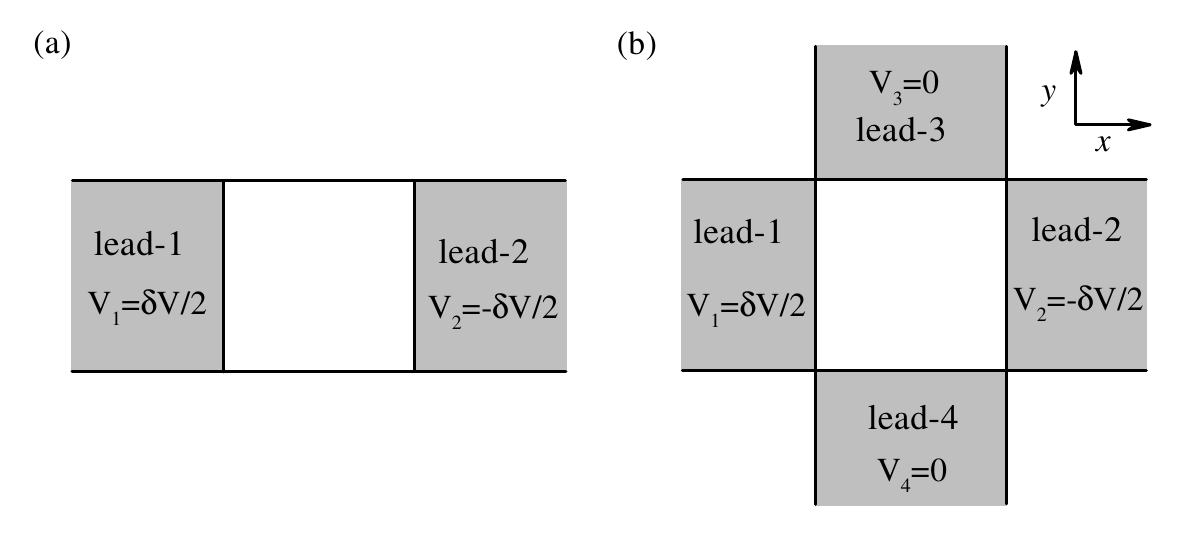}
\caption{Schematics of the two-dimensional system setups. (a) Two-terminal system: a central scattering region connected to two semi-infinite leads (labeled by lead-$1$ and lead-$2$). (b) Four-terminal system: a central scattering region connected to four leads labeled by lead-$\{1, 2, 3, 4\}$. A small bias $\delta V$ is symmetrically applied on two leads in the $x$ direction.}\label{fig:fig1}
\end{figure}

\subsection{Example: average second-order conductance in a four-terminal system}\label{sec:IIB}

In this subsection, we further demonstrate the versatility of our approach by studying the disorder-averaged second-order conductance in a four-terminal system.

The second-order conductance has been formulated using Green's functions in Refs.~[\onlinecite{Wang}] and [\onlinecite{Wei}],
\begin{equation}
\begin{aligned}
T_{\alpha \beta \gamma}= & -\frac{1}{2}\delta_{\beta \gamma} \operatorname{Tr}\left[\left(\Gamma \delta_{\alpha \gamma}-\Gamma_\gamma\right)\right. \\
& \left.\times\left(G^a \Gamma_\alpha G^r G^r+G^a G^a \Gamma_\alpha G^r\right)\right],
\end{aligned}
\end{equation}
where $\alpha, \beta, \gamma$ are lead indices and $\delta_{\beta\gamma}, \delta_{\alpha\gamma}$ are Kronecker delta functions. $\Gamma=\sum_{\alpha}\Gamma_{\alpha}$ is the total linewidth for all leads. Without loss of generality, we present the result of one second-order conductance $T_{311}$, where index $1, 2, 3, 4$ denotes the left, right, up and down leads, respectively [see Fig.~\ref{fig:fig1} (b)].

Expanding $G^{r}$ and keeping the fourth-order of disorder $V$, we get the average second-order conductance,
\begin{equation}\label{eq:<T311>}
\langle T_{311}\rangle = T_{311}^{(0)} + \langle T_{311}^{(2)}\rangle + \langle T_{311}^{(4)}\rangle + O(V^{6}).
\end{equation}
By defining $D_3=g^a\Gamma_3g^r$ and $D_1=g^r\Gamma_1g^a$, the zeroth-order, second-order and fourth-order terms are given by
\begin{equation}\label{eq:T3110}
T_{311}^{(0)}=\frac{1}{2} \operatorname{Tr}\left[\left(D_3 g^r+g^a D_3\right)\Gamma_1\right], \nonumber
\end{equation}
\begin{equation}\label{eq:T3112}
\begin{split}
\langle T_{311}^{(2)}\rangle=&\frac{1}{2}\operatorname{Tr}\Big[
  g^r \Gamma_1 D_3 \langle V g^r g^r V\rangle + g^r D_1 \langle V D_3 V\rangle\\
&+D_1 \langle V D_3 g^r V\rangle + D_3 D_1 \langle V g^a V\rangle\\
&+g^r g^r \Gamma_1 D_3 \langle V g^r V\rangle+g^r \Gamma_1 D_3 g^r \langle V g^r V\rangle\Big] + c.c.,
\end{split} \nonumber
\end{equation}
and
\begin{equation}\label{eq:T3114}
\begin{split}
\langle T_{311}^{(4)}\rangle=&\frac{1}{2}\operatorname{Tr}\Big[
   g^r \langle V_4^r\rangle g^r \Gamma_1 D_3 + g^r \langle V_3^r D_1 V\rangle D_3 \\
&+ g^r \langle V_3^r g^r \Gamma_1 D_3 V\rangle g^r + g^r \langle V_2^r D_1 V_2^a\rangle D_3 \\
&+ g^r \langle V_2^r D_1 V D_3 V\rangle g^r + g^r \langle V_2^r g^r \Gamma_1 D_3 V_2^r\rangle g^r \\
&+ g^r \langle V D_1 V_3^a \rangle D_3 + g^r \langle V D_1 V_2^a D_3 V\rangle g^r \\
&+ g^r \langle V D_1 V D_3 V_2^r\rangle g^r + g^r \langle V g^r \Gamma_1 D_3 V_3^r\rangle g^r \\
&+ D_1 \langle V_4^a\rangle D_3 + D_1 \langle V_3^a D_3 V\rangle g^r \\
&+ D_1 \langle V_2^a D_3 V_2^r\rangle g^r + D_1 \langle V D_3 V_3^r \rangle g^r \\
&+ g^r \Gamma_1 D_3 \langle V_4^r\rangle g^r
\Big] + c.c.,
\end{split} \nonumber
\end{equation}
where the notations $V_n^{r,a}$ are defined as $V_2^{r,a}=Vg^{r,a}V$ and $V_3^{r,a}=Vg^{r,a}Vg^{r,a}V$, for example.

By merging matrix productions between disorder matrix $V$ into single matrices, the above terms can be evaluated using the same method as presented in the previous subsection. This allows us to determine the coefficients $a_0$, $a_2$ and $a_{4}$ for the average second-order conductance,
\begin{equation}
\langle T_{311}\rangle = a_0 + a_2 W^2 + a_4 W^4 + O(W^6).
\end{equation}
This expression is similar to Eq.~(\ref{eq:Tw}), which further affirms that our approach is widely applicable, irrespective of specific models and physical quantities. We emphasize that our method is not limited to the fourth-order expansion; higher-order expansions can be incorporated in a similar manner to improve accuracy. For instance, in Appendix~\ref{appendix}, we present the sixth-order expansion in the calculation of Eq.~\eqref{eq:<T>}, demonstrating the method's flexibility and its potential for further refinement.

\section{NUMERICAL RESULTS}\label{sec:res}

In this section, we present the results of our method on three models and various physical quantities, and compare them with brute force calculations. We calculate the first-order and second-order conductances using the tight-binding method for a NM system, a spin Hall system with Rashba spin-orbital coupling (SOC) and a tilted Dirac model. The two-terminal and four-terminal system systems in two dimensions are schematically shown in Fig.~\ref{fig:fig1}. In this work, the central scattering regions in all setups are square lattice with the same size of $20a\times 20a$, where $a$ is the lattice constant. Anderson disorder is introduced in the central region for all numerical calculations, represented by a random on-site potential drawn from a uniform distribution within the range $\left[-W/2,W/2\right]$.

Note that when Anderson disorder is added in the central region, all odd-order terms of the disorder strength $W$ are canceled. Thus, the disorder average of an quantity simplifies to
\begin{equation}
\langle {\cal{O}} \rangle = a_0 + a_2W^2 + a_4W^4 + O(W^6).
\end{equation}
We calculate these coefficients $a_n$ in the following examples.

\subsection{Linear and second-order conductances in two-terminal system}\label{sec:NM}

We start by computing the average first and second-order conductances for a two-terminal normal metal (NM) system, as illustrated in Fig.~\ref{fig:fig1}(a). The hopping parameter is set to $t=1$, serving as the energy unit. The Fermi energy is $E_F=0.0526$, resulting in the opening of only one transmission channel.

In the presence of disorder, the average first-order conductance is suppressed upon increasing disorder strength, ultimately reaching zero when the transport is completely blocked by disorder. The average conductance $\langle T\rangle$ is calculated using Eq.~(\ref{eq:<TT>}). Following the application of our method in Sec.~\ref{sec:theory} A, we derive the disorder strength dependence of the average conductance, as expressed in Eq.~(\ref{eq:Tw}), with coefficients $a_0=1, a_2=-0.997$, and $a_4=1.797$.

\begin{figure}[tbp]
\includegraphics[width=8.5cm]{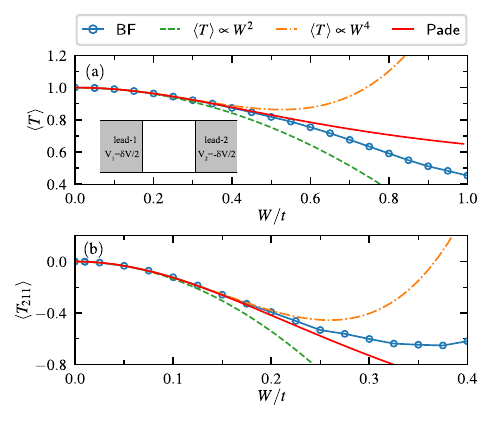}
\caption{(Color online) The disorder-averaged linear conductance (a) and the second-order conductance $\langle T_{211}\rangle$ (b) in a two-terminal NM system. Inset: schematic of the two-terminal device. Different lines correspond to different calculation methods.}\label{fig:fig2}
\end{figure}

The results of the average conductance obtained through different methods are presented in Fig.~\ref{fig:fig2}(a). We collected 10,000 disorder samples in BF calculation to obtain a smooth curve. The BF result encompasses all orders of the disorder matrix $V$, and hence serves as the reference curve for other methods. The results obtained through our method in the second- and fourth-order of $W$ show accurate predictions in the strength range up to $W=0.3$ and $0.4$, respectively. In the limit of $W\to 0$, $\langle T\rangle$ approaches $T_0$, which equals unity for only one transmission channel. As the disorder strength increases, our method in the second-order of $W$ provides a good approximation in the range $W<0.25$. By incorporating the fourth-order terms, this range expands to $W<0.4$. Nonetheless, higher-order terms such as 6$^{th}$- and 8$^{th}$-order terms can always be included using the same procedures to achieve more accurate approximation of $\langle T\rangle$. If we expand the fourth-order analytical result in Eq.~(\ref{eq:Tw}) using the Pad\'{e} approximation, we get a quite good estimation to the BF result. The Pad\'{e} approximation is given by
\begin{equation}\label{eq:pade}
\langle T(W)\rangle=\frac{\beta_1+\beta_2W^2}{\alpha_1+\alpha_2W^2},
\end{equation}
with $\alpha_1=4.36\times 10^{5}, \alpha_2=7.91\times 10^{5}, \beta_1=43.57, \beta_2=3.59$. This Pad\'{e} expansion is based on the analytic result in Eq.~(\ref{eq:Tw}), but it has higher accuracy and wider applicable range, which is widely used in conventional statistical analysis\cite{WangJ}.

In Fig.~\ref{fig:fig2}(b), we illustrate the performance of our method on the average second-order conductance. At the weak disorder limit, the second-order conductance is zero, which is expected for a trivial normal metallic system. The approximation $\langle T \rangle \propto W^2$ gives an accurate $\langle T_{211}\rangle$ for $W<0.1$. While including fourth-order terms, the range extends to $W<0.2$. The coefficients are given by $a_0=0, a_2=-13.524$, and $a_4=100.425$. The Pad\'{e} expansion Eq.~(\ref{eq:pade}) offers an overall better estimation with $\alpha_1=1.404\times10^8, \alpha_2=1.043\times10^9, \beta_1=1.328, \beta_{2}=1.899\times10^9$.

\subsection{Spin Hall conductance in four-terminal system}\label{sec:SH}

\begin{figure}
\includegraphics[width=\columnwidth]{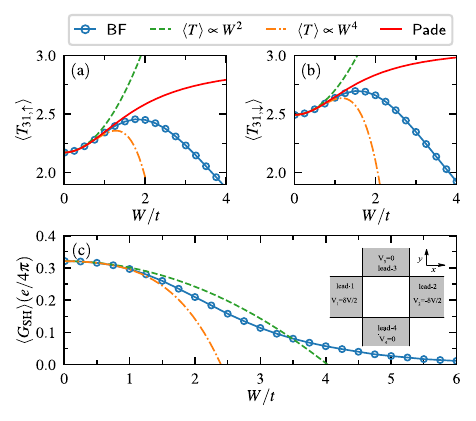}
\caption{(Color online) Disorder-averaged spin-resolved linear conductance $T_{31, \uparrow}$ (a) and $T_{31, \downarrow}$ (b) in a four-terminal system with Rashba SOC. (c) The average spin Hall conductance as a function of disorder strength $W$. Inset of (c): A schematic view of the four-terminal device. The Rashba SOC exists in the central region, lead-1 and lead-2. A small voltage bias $\delta V$ is applied across lead-1 and lead-2, while the spin Hall conductance is measured through leads 3 and 4. Parameters: $t_{\rm so}=0.5, \delta V=0.1$, Fermi energy $E_F=2$.}\label{fig:fig3}
\end{figure}

In this subsection, we study the spin Hall conductance in a four-terminal system along with the corresponding linear conductances, and compare our method with the BF results. The 2-dimensional four-terminal system is shown in the inset of Fig.~\ref{fig:fig3}(c). The tight-binding Hamiltonian with Rashba SOC is given by~\cite{Sheng,Ren,Qiao,Xing}
\begin{equation}
\begin{aligned}
H= & -t \sum_{\langle i j\rangle \sigma}( c_{i,\sigma}^{\dagger} c_{j,\sigma}+\text{H.c.})+\sum_{i \sigma} \varepsilon_i c_{i, \sigma}^{\dagger} c_{i, \sigma} \\
& -t_{\mathrm{so}} \sum_i\left[\left(c_{i,\uparrow}^{\dagger} c_{i+\delta_x, \downarrow}-c_{i, \downarrow}^{\dagger} c_{i+\delta_x, \uparrow}\right)\right. \\
& \left.-i\left(c_{i, \uparrow}^{\dagger} c_{i+\delta_y \downarrow}+c_{i, \downarrow}^{\dagger} c_{i+\delta_y, \uparrow}\right)+\text { H.c. }\right],
\end{aligned}
\end{equation}
where $t_{\rm so}$ is the effective coupling strength of the Rashba interaction. To ensure a well-defined measurement of spin current, we consider the case where SOC is present everywhere except in leads 3 and 4~\cite{Souma, Qiao, Ren}. In numerical calculation, we set $t_{\rm so}=0.5$, voltage bias $\delta V=0.1$, and Fermi energy $E_F=2$. The spin Hall conductance is defined as
\begin{equation}
G_{\rm SH}=(e/4\pi)\left(T_{31, \uparrow}-T_{31, \downarrow}\right),
\end{equation}
where $T_{31,\sigma}={\rm Tr}[\Gamma_{3\sigma}G^r\Gamma_1G^a]$ is the electron transmission coefficient from lead 1 to spin-$\sigma$ subband in lead 3, namely the spin-resolved linear conductance.

Fig.~\ref{fig:fig3}(a) and (b) show the spin-resolved average linear conductances $T_{31,\uparrow}$ and $T_{31,\downarrow}$. Our method in the fourth-order of $V$ agrees well with BF calculation for $W<1$. The coefficients for $T_{31,\uparrow}$ are $a_0=2.171, a_2=0.23, a_4=-0.0709$, while for $T_{31,\downarrow}, a_0=2.492, a_2=0.211, a_4=-0.0772$. The spin Hall conductance in Fig.~\ref{fig:fig3}(c) exhibits similar accuracy, with $a_0=0.321, a_2=-0.0198, a_4=-0.00628$.

\subsection{Second-order Hall effect in four-terminal system}\label{sec:Dirac}

\begin{figure}
\includegraphics[width=8.5cm]{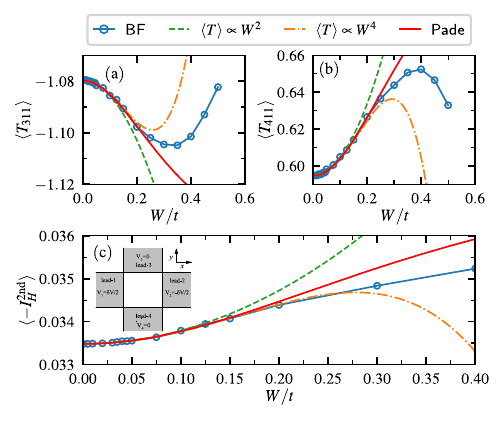}
\caption{(Color online) The average second-order conductances $T_{311}$ (a) and $T_{411}$ (b) for the tilted Dirac model in a four-terminal system. (c) The second-order Hall current as a function of $W$. Parameters: $B=1,\delta =-0.25,{{v}_{y}}=1.0, D=0.1, \delta V=0.2$.}\label{fig:fig4}
\end{figure}

In this subsection, we present the findings regarding the second-order conductance and second-order Hall current in a tilted Dirac model. To explore the second-order nonlinear Hall effect, we employ a Hamiltonian hosting tilted massive Dirac cones, which preserves time-reversal symmetry and breaks inversion symmetry~\cite{Papaj,Sodemann,Tanaka}. In the tight-binding representation, the Hamiltonian on a square lattice is expressed as\cite{Wei}
\begin{equation}\label{eq:Dc}
H=\sum_\mathbf{i}\left[ c_{\mathbf{i}}^{\dag}{{T}_{0}}{{c}_{\mathbf{i}}} + \left(c_\mathbf{i}^\dag T_x c_{\mathbf{i}+\mathbf{a}_x} + c_\mathbf{i}^\dag T_y c_{\mathbf{i}+\mathbf{a}_y} \right) + {\rm H.c.} \right], \nonumber
\end{equation}
with
\begin{eqnarray}
{{T}_{0}}& = &  -4 T_x +\delta {{\tau }_{z}}+D{{\tau }_{x}}, \nonumber \\
{T_x}    & = &  - {B{\tau _z}}/a^2, \nonumber \\
{T_y}    & = &  T_x - iv_y{\tau _y}/(2a). \nonumber
\end{eqnarray}
Here $c_{\mathbf{i}}^{\dag}$ denotes the creation operator on the discrete lattice $\mathbf{i}$, and $\tau_{x,y,z}$ denote the Pauli matrices. The Dirac cone is tilted due to $v_y$ term. The system is schematically depicted in the inset of Fig.~\ref{fig:fig4}(c), where four semi-infinite leads are connected to the central region. We calculate the second-order conductances and Hall current with a small bias applied across lead 1 and 2 in the $\hat x$ direction. In numerical calculations, Hamiltonian parameters are set as $B=1,\delta =-0.25,{{v}_{y}}=1.0$ and $D=0.1$. The voltage bias is $\delta V=0.2$. The Fermi energy $E_F=0.1225$.


The average second-order conductances are shown in Fig.~\ref{fig:fig4}(a) and (b). Notice that the BF result is averaged over 100,000 disorder samples, since the averages of higher-order quantities are more difficult to converge. Our method $\langle T \rangle \propto W^2$ shows a good agreement with BF in the range of $W<0.1$, and the range expands to $W<0.2$ if we include the fourth-order of $V$. The coefficients up to the fourth-order of $W$ is given by $a_0=-1.08, a_2=-0.577, a_4=4.341$ for $\langle T_{311}\rangle$, and $a_0=0.595, a_2=0.976, a_4=-5.705$ for $\langle T_{411}\rangle$. The Pad\'{e} approximation gives a good estimation for $W<0.25$.

It is well known that in this system, nonzero second-order nonlinear Hall response shows up while the linear Hall effect vanishes due to the presence of time-reversal symmetry~\cite{Sodemann}. We calculate the second-order Hall current in this system, which is expressed as\cite{Wei}
\begin{equation}
I_H^{\rm 2nd} = (T_{311}V_1^2 + T_{322}V_2^2) - (T_{411}V_1^2 + T_{422}V_2^2).
\end{equation}
Note that in this system only mirror symmetry $\mathcal{M}_y$ is broken by $v_y$, and $\mathcal{M}_x$ symmetry still exists. Therefore, we have $T_{311}=T_{322}$ and $T_{411}=T_{422}$~\cite{Wei}. The result of average second-order Hall current is shown in Fig.~\ref{fig:fig4}(c). The fourth-order approximation gives good estimation of $\langle I_H^{\rm 2nd}\rangle$ for $W<0.25$ with $a_0=0.0335, a_2=0.031, a_4=-0.201$. Fig.~\ref{fig:fig4}(c) clearly shows enhancement behavior of the second-order Hall current upon increasing of the disorder strength. This numerical result confirms previous theoretical prediction on disorder enhancement of the second-order Hall effect in bulk systems~\cite{Du2019,Du2021}. This work, together with Ref.~[\onlinecite{Wei2022}], provide numerical evidence for disorder enhancement of the second-order and third-order Hall effect in four-terminal systems in quantum transport regime.

\begin{table}
\caption{\label{tab:1} Coefficients of various quantities obtained by our method for different models. Models A, B and C refers to the three model systems in Sec.~\ref{sec:NM},~\ref{sec:SH} and~\ref{sec:Dirac}, respectively.}
\begin{ruledtabular}
\begin{tabular}{cccccc}
Model & quantity & $a_{0}$ & $a_{2}$ & $a_4$ & $W_{\max}/t$ \\ \toprule
A & $\langle T\rangle$ & 1 & -0.997 & 1.797 & 0.4\\
A & $\langle T_{211}\rangle$ & 0 & -13.524 & 100.425 & 0.2\\
B & $\langle T_{31,\uparrow}\rangle$ & 2.171 & 0.23 & -0.0709 & 1\\
B & $\langle T_{31,\downarrow}\rangle$ & 2.492 & 0.211 & -0.0772 & 1\\
B & $\langle G_{\rm SH}\rangle$ & 0.321 & -0.0198 & -0.00628 & 1\\
C & $\langle T_{311}\rangle$ & -1.08 & -0.577 & 4.341 & 0.2\\
C & $\langle T_{411}\rangle$ & 0.595 & 0.976 & -5.705 & 0.2\\
C & $\langle -I_{\rm H}^{\rm 2nd}\rangle$ & 0.0335 & 0.031 & 0.201 & 0.25\\
\end{tabular}
\end{ruledtabular}
\end{table}

The coefficients of various quantities for each model are summarized in Table~\ref{tab:1} to facilitate quick reference. Here, $W_{\max}$ is the maximum disorder strength blow which $\langle {\cal{O}} \rangle=a_0+a_2W^2+a_4W^4$ gives accurate prediction of the average of quantity ${\cal{O}}$. Based on this table, we have the following observations. (1) The method works better for linear conductances than for second-order conductances. This is the direct consequence of perturbation expansion. (2) For linear conductance, the method is more accurate in four-terminal systems than in two-terminal ones. (3) For the second-order conductance, the method gives nearly the same $W_{\max}$ in two-terminal and four-terminal systems. We also mention that it is very difficult to obtain converged disorder averages in theses models using the BF method. The typical disorder samples needed ranges from 10,000 to 100,000. In contrast, our method only requires a single calculation to obtain the analytical expression for approximating the disorder average. Moreover, the complexity brought in by high-order truncation can be handled in a recursive way, which can be easily realized in codes.

\section{CONCLUSION}\label{sec:conclusion}
In this work, we have developed a general method for calculating average transport properties in disordered noninteracting systems based on the nonequilibrium Green's function formalism. This method extracts the coefficients of average quantities in orders of the disorder strength. The only approximation involved is the truncation in the Dyson equation, which can always be improved by including high-order terms. Moreover, the accuracy of this method can be explicitly controlled. The difference between the averages of two consecutive orders in the expansion can serve as a criterion for determining accuracy. If the difference falls below a specified threshold, the expansion can be truncated at that order.

To demonstrate the applicability of this method, we calculated the average linear and second-order nonlinear conductances for three different models. We found that truncation in the fourth order already gives accurate estimations on various average quantities in a wide range of disorder. Our method gives the analytical expressions of disorder averages as a function of the disorder strength, which avoids the time-consuming BF calculations. The coefficients in the analytical expressions for various average properties in different models are summarized in Table~\ref{tab:1}. Considering the limitations of existing methods, our general method provides a powerful tool for disorder-averaged calculation in quantum transport regime.

\section*{ACKNOWLEDGMENTS}
This work was supported by the National Natural Science Foundation of China (Grants No. 12034014 and No. 12174262).

\section*{APPENDIX}\label{appendix}
In this appendix, we derive the sixth-order terms in Eq.~\eqref{eq:<T>} as an exmaple of high-order calculations. Expanding Green's functions in Eq.~\eqref{eq:<TT>} using Dyson equation, we get the sixth-order term
\begin{equation}\label{T6}
\begin{split}
\langle T^{(6)}\rangle =& {\rm Tr} [D_L \Gamma_R g^a\langle Vg^aVg^aVg^aVg^aVg^aV\rangle] \\
& + {\rm Tr}[D_L\langle Vg^r\Gamma_Rg^aVg^aVg^aVg^aVg^aV\rangle] \\
& + {\rm Tr}[D_L\langle Vg^rVg^r\Gamma_Rg^aVg^aVg^aVg^aV\rangle] \\
& + \frac{1}{2}{\rm Tr}[D_L\langle Vg^rVg^rVg^r\Gamma_Rg^aVg^aVg^aV\rangle] + c.c.
,
\end{split}
\end{equation}
where matrix $D_L$ is defined in the main text. Following the same procedure in main text, we write each term above as
\begin{equation}
\langle T_i^{(6)} \rangle = {\rm Tr}[A\langle VBVCVDVEVFV\rangle]
\end{equation}
with $A, B, C, D, E$ and $F$ the matrix productions independent of disorder matirx $V$. Defining matrix $X_{ij}^{(6)}=\sum_{klmn} B_{ik}C_{kl}D_{lm}E_{mn}F_{nj}\epsilon_{ijklmn}$, we have
\begin{equation}
\epsilon_{ijlkmn} = \langle V_{ii}V_{kk}V_{ll}V_{mm}V_{nn}V_{jj}\rangle.
\end{equation}
This average can be decomposed into $\langle V_{ii}^6\rangle$ and lower-order terms $\langle V_{ii}^4\rangle$ and $\langle V_{ii}^2\rangle$. Lower-order terms are given previously, the sixth-order term is given by
\begin{equation}
\langle V_{ii}^6\rangle = \frac{1}{W}\int_{-W/2}^{W/2} V^6 dV = \frac{W^6}{448}.
\end{equation}
For example, for diagonal elements $X_{ij}^{(6)}$ with $i=j$, we have
\begin{equation}
\begin{split}
\epsilon_{ijlkmn} =& W^6\delta_{ij}\Big[\frac{\delta_{jk}\delta_{kl}\delta_{lm}\delta_{mn}}{448} + \frac{1}{960}(\delta_{kl}\delta_{lm}\delta_{mn} \\
& + \delta_{jk}\delta_{kl}\delta_{mn} + \delta_{jk}\delta_{km}\delta_{ln} + \delta_{jk}\delta_{kn}\delta_{ml} \\
&+ \delta_{jl}\delta_{lm}\delta_{kn} + \delta_{jl}\delta_{ln}\delta_{mk} + \delta_{jm}\delta_{mn}\delta_{kl}) \\
& + \frac{1}{1728}(\delta_{kl}\delta_{mn} + \delta_{km}\delta_{ln} + \delta_{kn}\delta_{ml} )\Big].
\end{split}
\end{equation}
Similarly, the factor $\epsilon_{ijkkmn}$ for $i\neq j$ can be obtained. After obtaining each term in Eq.~\eqref{T6}, we then get the sixth-order term $\langle T^{(6)}\rangle$ and the corresponding coefficient $a_6$.


\begin{thebibliography}{00}
\bibitem{note1}
The recursive Green's function method is suitable for both clean and disordered systems.
\bibitem{note2}
In this work, we use Anderson-tpye disorder to illustrate our method. For other diagonal-disorder, Eqs.~(\ref{eq:V2}) and (\ref{eq:V4}) become $\langle V_{ii}V_{jj}\rangle = \frac{1}{W} \int_{-W/2}^{W/2} {\rm d}V\rho(V)V^2 \delta_{ij}$ and $\langle V_{ii}V_{jj}V_{kk}V_{ll}\rangle = \frac{1}{W} \int_{-W/2}^{W/2} {\rm d}V\rho(V)V^4 \delta_{ij}\delta_{jk}\delta_{kl}$, respectively. Here $\rho(V)$ is the probability density function of that disorder. While for off-diagonal disorder, our method is limited to being applicable only to some simple disorder types.
\end{thebibliography}
\end{document}